\documentclass{elsart}
\usepackage{graphicx}
\usepackage{color}
\usepackage{amssymb}
\usepackage[all]{xy}

\newcommand{\CSP}[0]{\ensuremath{\textsc{CSP}}}
\newcommand{\QCSP}[0]{\ensuremath{\textsc{QCSP}}}
\newcommand{\CSPc}[0]{\ensuremath{\textsc{CSP}_c}}
\newcommand{\QCSPc}[0]{\ensuremath{\textsc{QCSP}_c}}
\renewcommand{\L}[0]{\ensuremath{\mathsf{L}}}
\newcommand{\Ptime}[0]{\ensuremath{\mathsf{P}}}
\newcommand{\Pspace}[0]{\ensuremath{\mathsf{Pspace}}}
\newcommand{\NP}[0]{\ensuremath{\mathsf{NP}}}
\newcommand{\ALogtime}[0]{\ensuremath{\mathsf{ALogtime}}}
\begin{document}

\begin{frontmatter}

\title{Low-level dichotomy for Quantified Constraint Satisfaction Problems}
\author{Barnaby Martin\thanksref{epsrc}}

\address{
School of Engineering and Computing Sciences, Durham University,\\
  Science Labs, South Road, Durham DH1 3LE, U.K.}

\thanks[epsrc]{Supported by EPSRC grant EP/G020604/1.}

\begin{abstract}
Building on a result of Larose and Tesson for constraint satisfaction problems (\CSP s), we uncover a dichotomy for the quantified constraint satisfaction problem $\QCSP(\mathcal{B})$, where $\mathcal{B}$ is a finite structure that is a core. Specifically, such problems are either in \ALogtime\ or are \L-hard. This involves demonstrating that if $\CSP(\mathcal{B})$ is first-order expressible, and $\mathcal{B}$ is a core, then $\QCSP(\mathcal{B})$ is in \ALogtime.

We show that the class of $\mathcal{B}$ such that $\CSP(\mathcal{B})$ is first-order expressible (indeed, trivially true) is a microcosm for all \QCSP s. Specifically, for any $\mathcal{B}$ there exists a $\mathcal{C}$ such that $\CSP(\mathcal{C})$ is trivially true, yet $\QCSP(\mathcal{B})$ and $\QCSP(\mathcal{C})$ are equivalent under logspace reductions.
\end{abstract}

\end{frontmatter}

\section{Introduction}
The \emph{constraint satisfaction problem} $\CSP(\mathcal{B})$, much studied in artificial intelligence, is known to admit several equivalent formulations, two of the best known of which are the query evaluation of primitive positive sentences -- those involving only existential quantification and conjunction -- on $\mathcal{B}$, and the homomorphism problem to $\mathcal{B}$ (see, e.g., \cite{KolaitisVardiBook05}). $\CSP(\mathcal{B})$ is \NP-complete in general, and a great deal of effort has been expended in classifying its complexity for certain restricted cases. The problems $\CSP(\mathcal{B})$ are conjectured \cite{FederVardi,JBK} to be always in \Ptime\ or \NP-complete. While this has not been settled in general, a number of partial results are known -- e.g. over structures of size $\leq 3$ \cite{Schaefer,BulatovJACM} and over smooth digraphs \cite{HellNesetril,SmoothDigraphs}. The classification project is simplified by the fact that one only needs to consider $\mathcal{B}$ that are \emph{cores} (for which every endomorphism is an automorphism), as it is is easy to see that for each $\mathcal{B}$ there is an induced substructure $\mathcal{B}'\subseteq \mathcal{B}$ such that $\CSP(\mathcal{B}')$=$\CSP(\mathcal{B})$ and $\mathcal{B}'$ is a core.

A popular generalisation of the CSP involves considering the query evaluation problem for \emph{positive Horn} logic -- those involving only the two quantifiers together with conjunction. The ensuing \emph{quantified constraint satisfaction problems} $\QCSP(\mathcal{B})$  allow for a broader class, used in artificial intelligence to capture non-monotonic reasoning, whose complexities rise to \Pspace. The only known complexities for $\QCSP(\mathcal{B})$ are \Ptime, \NP-complete and \Pspace-complete -- though a trichotomy is not yet formally conjectured. For results on the complexity of QCSPs, see \cite{Chenetal,HubieSicomp}; for partial trichotomies, see \cite{Chenetal,CiE2006}.

While the greatest endeavour of the CSP classification project has been towards the separation of these problems between \Ptime\ and \NP-complete, a recent result -- derived through the algebraic method -- gave a much lower-level complexity dichotomy.
\begin{thm}[\cite{LaroseTesson2009}]
For finite $\mathcal{B}$, either $\CSP(\mathcal{B})$ is first-order expressible, or $\CSP(\Gamma)$ is \L-hard under first-order reductions.
\end{thm}
\noindent By the well known containment of first-order evaluation in alternating logarithmic time (\ALogtime) -- see, e.g. \cite{Immerman} -- this gives the following complexity-theoretic dichotomy.
\begin{cor}
For finite $\mathcal{B}$, either $\CSP(\mathcal{B})$ is in \ALogtime, or $\CSP(\mathcal{B})$ is \L-hard under first-order reductions.
\end{cor}
In our first result of this note, Corollary~\ref{cor:main1}, we uncover a similar dichotomy for QCSP, in the case where $\mathcal{B}$ is a core. Since there is a trivial reduction from $\CSP(\mathcal{B})$ to $\QCSP(\mathcal{B})$, what remains is to demonstrate that $\QCSP(\mathcal{B})$ is in \ALogtime, when $\CSP(\mathcal{B})$ is first-order expressible and $\mathcal{B}$ is a core. 

It is easy to see that the \QCSP\ can not have a dichotomy between \ALogtime\ and \L-hard at the same position as the \CSP\ as there is even a Boolean $\mathcal{B}$ such that $\CSP(\mathcal{B})$ is first-order expressible while $\QCSP(\mathcal{B})$ is \Pspace-complete (take a $0$-valid $\mathcal{B}$ that is not Horn, dual Horn, affine or bijunctive -- see \cite{Nadia}). For our second result, Proposition~\ref{prop:main2}, we are able to go further and state that the class of $\mathcal{B}$ such that $\CSP(\mathcal{B})$ is first-order expressible (indeed, trivially true) is a microcosm for all QCSPs, in the following sense. For any $\mathcal{B}$ there exists a $\mathcal{C}$ such that $\CSP(\mathcal{C})$ is trivially true, yet $\QCSP(\mathcal{B})$ and $\QCSP(\mathcal{C})$ are equivalent under logspace reductions.

\section{Preliminaries}

Let $\mathcal{B}$ be a finite structure over signature $\sigma$ whose domain is $B$. The \emph{quantified constraint satisfaction problem} $\QCSP(\mathcal{B})$ takes as an input a sentence $\Phi$ of the form $\forall u_1 \exists v_1,\ldots,\forall u_n \exists v_n \ \phi(u_1,v_1,\ldots,u_k,v_k)$, where $\phi$ is a conjunction of atoms (such a sentence is termed \emph{positive Horn}). 
The input is a yes-instance iff $\mathcal{B} \models \Phi$. We denote yes-instances by $\Phi \in \QCSP(\mathcal{B})$. The \emph{constraint satisfaction problem} $\CSP(\mathcal{B})$ is similarly defined, but with instances $\Phi$ of the form $\exists v_1,\ldots,v_n \ \phi(v_1,\ldots,v_n)$ (such a sentence is termed \emph{primitive positive} -- by convention the Boolean $\bot$ is also considered to be primitive positive and positive Horn). Finally, the problems $\QCSPc(\mathcal{B})$ and $\CSPc(\mathcal{B})$ are defined as $\QCSP(\mathcal{B})$ and $\CSP(\mathcal{B})$, respectively, but where we allow constants of $\mathcal{B}$ to appear in the instances. A sentence $\Phi$ that is an instance of $\CSP(\mathcal{B})$ naturally defines a structure $\mathcal{D}_\Phi$, termed the \emph{canonical database} of $\Phi$, over the same signature as $\mathcal{B}$. The domain of $\mathcal{D}_\Phi$ consists of the variables of $\Phi$ and the relations consist of those relations listed in the body $\phi$ of $\Phi$. Conversely, for a structure $\mathcal{B}$, the existential quantification of the conjunction of the facts (relation tuples) of $\mathcal{B}$, $\Phi_\mathcal{B}$, is the \emph{canonical query} of $\mathcal{B}$ (see, e.g., \cite{KolaitisVardiBook05}). A \emph{homomorphism} from $\mathcal{B}$ to $\mathcal{C}$, structures over the same signature, is a function $f$ from $B$ to $C$ such that whenever $(x_1,\ldots,x_r) \in R^\mathcal{B}$, then $(f(x_1),\ldots,f(x_r)) \in R^\mathcal{C}$. An \emph{endomorphism} of $\mathcal{B}$ is a homomorphism from $\mathcal{B}$ to itself. A \emph{core} is a finite structure $\mathcal{B}$, all of whose endomorphisms are automorphisms. It is known that, for every structure $\mathcal{B}$, there exists a unique (up to isomorphism) core $\mathcal{C}$ s.t. $\CSP(\mathcal{B})=\CSP(\mathcal{C})$. The relationship between canonical database and canonical query exposes the alternative guise of the \CSP\ as the \emph{homomorphism problem}, in which one observes that a primitive positive $\Phi$ is such that $\Phi \in \CSP(\mathcal{B})$ iff there is a homomorphism from $\mathcal{D}_\Phi$ to $\mathcal{B}$. We will often move implicitly between these two versions of the problem.

The problem $\CSP(\mathcal{B})$ is \emph{first-order expressible} if there is a first-order sentence $\Theta$ s.t. $\Phi \in \CSP(\mathcal{B})$ iff $\mathcal{D}_\Phi \models \Theta$. We will need the following result from \cite{LaroseLotenTardifJournal} (for the definition of, and more on, polymorphisms see \cite{CSPSurvey} or \cite{JBK}).
\begin{lem}[\cite{LaroseLotenTardifJournal}]
\label{lem:LLT}
If $\CSP(\mathcal{B})$ is first-order expressible and $\mathcal{B}$ is a core, then $\mathcal{B}$ has a $k$-ary near-unanimity polymorphism, for some $k$.
\end{lem}
We borrow the following definitions and results from \cite{HubieSicomp}. For $a \in B$, an instance $\Phi'$ of $\QCSPc(\mathcal{B})$ is a \emph{$(j,a)$-collapsing} of an instance $\Phi$ if $\Phi'$ is obtained from $\Phi$ by instantiating all but $j$ of the universal variables of $\Phi$ by $a$. We refer to the (at most $j$) universal variables that are not instantiated with $a$ as having \emph{survived} the collapsing. For $a \in B$, the structure $\mathcal{B}$ is said to be \emph{$(j,a)$-collapsible} when the following holds: iff all $(j,a)$-collapsings of an instance $\Phi$ of $\QCSPc(\mathcal{B})$ are true, then the instance $\Phi$ is true. 
\begin{thm}[\cite{HubieSicomp}]
\label{thm:hubie}
If $\mathcal{B}$ has a $k$-ary near-unanimity polymorphism then $\mathcal{B}$ is $(k-1,a)$-collapsible for all $a \in B$.
\end{thm}
\noindent We will make use of the construction used in the proof of the following lemma from \cite{HubieSicomp} -- for pedagogy we retain the original paper's notational format.
\begin{lem}[\cite{HubieSicomp}]
\label{lem:hubie}
Given a $(j,a)$-collapsing $\Phi'$ of an instance $\Phi$ of $\QCSP(\mathcal{B})$, there is an instance $\Phi''$ of $\CSPc(\mathcal{B})$ s.t. $\Phi'' \in \CSPc(\mathcal{B})$ iff $\Phi' \in \QCSPc(\mathcal{B})$.\footnote{Note the absence of constants from $\Phi$, that may be present in both $\Phi'$ and $\Phi''$.} 
\end{lem}
\begin{pf}[Sketch Proof]
The variables of $\Phi''$ are pairs of the form $(v_i,\alpha)$ where $\alpha$ is an assignment of elements of $\mathcal{B}$ to the (at most $j$) universal variables among $u_1,\ldots,u_{i-1}$ that survive the collapsing to $\Phi'$.

For each relation $R(x_1,\ldots,x_r)$ in $\Phi'$, and for each $\alpha$, we add the relations $R(w_1,\ldots,w_r)$ to $\Phi''$ where $w_i:=a$ if $x_i=a$, $w_i:=\alpha(u_l)$ if $x_i=u_l$, and $w_i:=(v_l,\alpha)$ if $x_i=v_l$.

From $\Phi'' \in \CSPc(\mathcal{B})$ we can read off the necessary alternating assignments that witness $\Phi' \in \QCSPc(\mathcal{B})$ and vice-versa. The result follows.
\end{pf}
\noindent In \cite{HubieSicomp}, essential use is made of the fact that the construction in the previous lemma of $\Phi''$ from $\Phi'$ may be accomplished in polynomial time. We will not have polynomial time to construct $\Phi''$, but we will be able to query its underlying structure (in \ALogtime). Observe that $\Phi''$ and $\Phi'$ may contain elements of $\mathcal{B}$ as constants, while $\Phi$ does not. The sound interpretation of constants through variables is the reason why our result applies only to the case of cores.

$\ALogtime$ is the class of all languages that can be decided by an alternating random access Turing Machine in logarithmic time. A random access Turing Machine may be considered to have two tapes, the first of which is only used to allow it to write the number of a square of the second to which the read/write head of the second may jump. One may imagine that the read/write head of the second tape jumps independently of the read/write head of the first (but it is easy to see that this is not important). For more on the diminutive class \ALogtime, see \cite{Immerman}. We will benefit from \emph{briefly} recalling how a first-order sentence $\Theta$ may be evaluated on a structure $\mathcal{D}$ in \ALogtime. Naturally, we may assume $\Theta$ is in prenex form, thus we guess universally and existentially through the, respectively, universal and existental quantifiers of $\Theta$. Finally, we have a quantifier-free statement involving a constant number of atomic queries which we may look up in the coding of $\mathcal{D}$. For example, if the structure has $n$ elements and we wish to know if $R(1,3,2)$ holds, then we jump to the $(1.n^2+3.n+2)$th bit in the encoding of $R$ to see if this is a $1$, asserting the tuple is in the relation. The number of steps this takes is clearly $O(|\Theta|.\log n)$ where $|\Theta|$ is the constant size of $\Theta$. 

Suppose now that $\mathcal{D}$ is presented as $\Phi_\mathcal{D}$, an input to $\CSP(\mathcal{B})$. Now $\mathcal{D} \models \Theta$ iff  $\Phi_\mathcal{D} \in \CSP(\mathcal{B})$ (recall that $\Theta$ expresses $\CSP(\mathcal{B})$). We may evaluate $\Theta$ directly on $\Phi_\mathcal{D}$ in the same manner as before, until we get to the quantifier-free query. But, though we know exactly where to jump in the coding of $\mathcal{D}$ to verify a particular relation tuple, we do not necessarily know where to jump in the coding of $\Phi_\mathcal{D}$. However, we can get around this in the following manner. If we need to verify a positive atom, then we existentially guess where it sits and jump there to verify the tuple's presence. If we need to verify a negative atom, then we universally guess all positions and verify the tuple's absence.

\section{Our first result}

Let $a_1,\ldots,a_m$ be an enumeration of the domain $B$ of $\mathcal{B}$, which will be a core, and let $\Phi$ be an instance of $\QCSP(\mathcal{B})$ of the form $\forall u_1 \exists v_1,\ldots,\forall u_n \exists v_n$ $\phi(u_1,v_1,\ldots,u_n,v_n)$. Let $\Theta$ be a first-order sentence expressing $\CSP(\mathcal{B})$ and let $k$ be s.t. $\mathcal{B}$ is $k$-collapsible (as guaranteed by Lemma~\ref{lem:LLT} and Theorem~\ref{thm:hubie}). We will specify structures $\mathcal{D}(\lambda_1,\ldots,\lambda_n)$, for $\lambda_1,\ldots,\lambda_k$ distinct in $[n]$, s.t. $\mathcal{B} \models \Phi$ iff, for all $\lambda_1,\ldots,\lambda_k$ distinct in $[n]$, $\mathcal{D}(\lambda_1,\ldots,\lambda_k) \models \Theta$ (iff, for all $\lambda_1,\ldots,\lambda_k$ distinct in $[n]$, $\mathcal{B} \models \Phi_{\mathcal{D}(\lambda_1,\ldots,\lambda_k)}$).

\

\noindent \textbf{Construction of $\mathcal{D}:=\mathcal{D}(\lambda_1,\ldots,\lambda_k)$}. Mostly this construction is as in Lemma \ref{lem:hubie}, only we will specify a structure and not a sentence, and we have to deal with (remove) the constants. Let $\Phi'$ be the $(k,a_1)$-collapsing of $\Phi$ given be the choice of universal variables surviving being $x_{\lambda_1},\ldots,x_{\lambda_k}$. The elements of the structure will be the elements $a_1,\ldots,a_m$ of $\mathcal{B}$ together with pairs of the form $(v_i,\alpha)$ where $\alpha$ is an assignment of elements of $\mathcal{B}$ to the (at most $k$) universal variables among $u_1,\ldots,u_{i-1}$ that survive the collapsing to $\Phi'$. For each relation tuple $R(x_1,\ldots,x_r)$ in $\Phi'$ we add the tuple $R(w_1,\ldots,w_r)$ to $\mathcal{D}$ where $w_i:=a_1$ if $x_i=a_1$, $w_i:=\alpha(u_l)$ if $x_i=u_l$, and $w_i:=(v_l,\alpha)$ if $x_i=v_l$. Finally, we add to $\mathcal{D}$ the relations on the constants -- all tuples $R(a_{\mu_1},\ldots,a_{\mu_r})$ that hold in $\mathcal{B}$. That $\mathcal{B} \models \Phi$ iff, for all $\lambda_1,\ldots,\lambda_k$ distinct in $[n]$, $\mathcal{D}(\lambda_1,\ldots,\lambda_k) \models \Theta$, follows as in the proof of Lemma~\ref{lem:hubie} from our definitions together with the fact that $\mathcal{B}$ is a core -- so in $\Phi_{\mathcal{D}(\lambda_1,\ldots,\lambda_k)}$ the variables associated to the constants $\{a_1,\ldots,a_m\}$, that make up the domain of $\mathcal{B}$, must evaluate to themselves in $\mathcal{B}$ (up to an isomorphism of their constraints -- see \cite{JBK}).

\

\noindent \textbf{Verification in \ALogtime}. Firstly, we must ensure that the instance is of the correct form -- something we could probably not do in, say, non-deterministic logarithmic time. If we did not do this, we could not be sure that the input was not a mix of rubbish that happened to contain somewhere the correct tuples. We check that the sentence begins $\forall u_1 \exists v_1$. We then existentially guess a point $\rho$ in the input and check that before $\rho$ comes $\forall u_n \exists v_n$ (for some $n$ -- note that the binary encoding of $n$ is of size $\log n$, hence we can read it and write it down). We then use universal guessing to check that everything before $\rho$ is of the local form $\forall u_i \exists v_i \forall u_{i+1} \exists v_{i+1}$, and that everything after $\rho$ consists locally of atoms of $\mathcal{B}$ none of whose variables is outside $u_1,v_1,\ldots,u_n,v_n$. It is clear this can be done in time $O(\log n)$.

Now we will attempt to verify the collapsings of the sentence. We universally guess $\lambda_1,\ldots,\lambda_k \in [n]$. We will now try to evaluate the first-order query $\Theta$ on $\mathcal{D}(\lambda_1,\ldots,\lambda_k)$, while the Turing Machine itself sees only the sentence $\Phi$. Let $\Phi'$ be the $(k,a_1)$-collapsing of $\Phi'$ derived from $\Phi$ by preserving the universal variables $u_{\lambda_1},\ldots,u_{\lambda_k}$. We may assume that $\Theta$ is in prenex normal form, thus we alternately universally and existentially guess the variables of $\Theta$ over the universe given by $\{a_1,\ldots,a_m\}$ $\cup$ $\{(v_i,\alpha): $ $v_i$ in $\Phi$ and $\alpha$ an assignment to the variables among $u_1,\ldots,u_{i-1}$ that survive the collapsing$\}$. Finally, it is necessary to evaluate relations from the quantifier-free part of $\Theta$. We first give the procedure for positive atoms. If the atom $R(w_1,\ldots,w_r)$, with $w_1,\ldots,w_r$ from $\mathcal{D}(\lambda_1,\ldots,\lambda_k)$ contains only  instances of $a_1,\ldots,a_m$ then we may first ascertain whether it is there due to being a tuple from $\mathcal{B}$. Otherwise, and if it contains instances of $a_1$, we must existentially guess how many of these came from instantiations to the universal variables that did not survive the collapsing and then existentially guess which variables they came from. Having done that, we existentially guess a point in the body $\phi$ of $\Phi$ and look for a relation tuple that could have given rise to $R(w_1,\ldots,w_r)$ under the substitution of $a_1$ to the guessed universal variables that did not survive the collapsing and something to the universal variables that did survive the collapsing that is consistent with the $\alpha$ of the $(v_i,\alpha)$ of the existential variables of $\Phi$. The procedure for negated atoms is dual. We universally guess that there is no position in the encoding and no instantiations of $a_1$ to any variables that did not survive the collapsing such that the relation tuple arises.\footnote{In fact the procedure only needs to deal with negated atoms. If $\CSP(\mathcal{B})$ is first-order expressible, it is in fact expressible in its universal-negative fragment \cite{DBLP:journals/ejc/Atserias08}.} Because the sentence is of constant size, it is not hard to see that this can be done in logarithmic time.

\begin{thm}
If $\CSP(\mathcal{B})$ is first-order expressible and $\mathcal{B}$ is a core then $\QCSP(\mathcal{B})$ is in \ALogtime.
\end{thm}
\begin{cor}
\label{cor:main1}
For finite cores $\mathcal{B}$, either $\QCSP(\mathcal{B})$ is in \ALogtime, or $\QCSP(\mathcal{B})$ is \L-hard under first-order reductions.
\end{cor}

\section{Our second result}

We wish to consider a positive Horn $\sigma$-sentence $\Phi:=\forall u_1 \exists v_1 \ldots \forall u_n \exists v_n \ \phi(u_1,$ $v_1,\ldots,u_n,v_n)$ as a \emph{partitioned $\sigma$-structure} $\mathfrak{P}_\Phi$ over $\sigma \uplus \{A_1,E_1,\ldots,A_n,E_n\}$, where $A_1,E_1,\ldots,A_n,E_n$ are unary relations that are either singleton or empty. The domain $P_\Phi$ of $\mathfrak{P}_\Phi$ consists of the variables of $\Phi$. The $\sigma$-relations of $\mathfrak{P}_\Phi$ are the atoms listed in $\phi$ and the unary relations $A_1,E_1,\ldots,A_n,E_n$ partition the elements according to where in the quantifer prefix $\forall u_1 \exists v_1 \ldots \forall u_n \exists v_n$ the corresponding variables sat. Conversely, from a $\sigma \uplus \{A_1,E_1,\ldots,A_n,E_n\}$ structure $\mathfrak{P}$, in which $A_n,E_n,\ldots,A_n,E_n$ partition the domain and have at most a single element in each, one may read a positive Horn $\sigma$-sentence $\Phi_\mathfrak{P}$ in the obvious fashion. The relationship between $\mathfrak{P}_\Phi$ and $\Phi$ is analogous to that between canonical database and canonical query (see \cite{CiE2006} for more on this construction). We will feel free to move between positive Horn $\Phi$ and partitioned structure $\mathfrak{P}$ -- in the latter we set $\mathcal{E}_{P}:= \bigcup_{i \in [n]} E_i$ and $\mathcal{A}_{P}:= \bigcup_{i \in [n]} A_i$.
\begin{prop}
\label{prop:main2}
Let $\mathcal{B}$ be a $\sigma$-structure on domain $B$. Then there is a binary relation $F$ and a $\sigma \uplus \{F\}$-structure $\mathcal{C}$, on domain $B \uplus \{c\}$, such that $\mathcal{C}$ is $c$-valid and $\QCSP(\mathcal{B})$ and $\QCSP(\mathcal{C})$ are logspace equivalent.
\end{prop}
\begin{pf}
We begin with the construction of $\mathcal{C}$. Recall $C:=B \uplus \{c\}$. For each $m$-ary relation $R \in \sigma$ we define 
\[R^\mathcal{C}:= R^\mathcal{B} \cup \{ (x_1,\ldots,x_m) : x_1,\ldots,x_m \in C \mbox{ and at least one $x_i$ is $c$} \}. \]
Finally, we define $F^\mathcal{C}$ to be $\{(c,c)\} \cup \{ (c,x) : x \in B \} \cup \{ (x,x') : x,x' \in B\}$ which can alternatively be given as $C^2 \setminus \{(x,c) : x \in B\}$.

($\QCSP(\mathcal{B}) \leq_\mathsf{L} \QCSP(\mathcal{C})$.) Given an input $\Phi$ for $\QCSP(\mathcal{B})$ we build an input $\Psi$ for $\QCSP(\mathcal{C})$ in the following fashion. From the partitioned $\sigma$-structure $\mathfrak{P}_\Phi$ build the partitioned $\sigma \uplus \{F\}$-structure $\mathfrak{P}'$ as follows. Firstly, we add two new partitions $A_0$ and $E_0$ and a new element $a \in A_0$ ($E_0$ will be empty).\footnote{Deviation from the previously specified indexing of $A_i,E_i$ should cause no trouble.} We then add the following directed $F$-edges: $\{(a,y): y\in \mathcal{E}_{P_\Phi}\}$,   $\{(y,y'):\mbox{$y,y' \in \mathcal{E}_{P_\Phi}$}\}$ and $\{(x,y):\mbox{$x \in \mathcal{A}_{P_\Phi}$ and $y \in \mathcal{E}_{P_\Phi}$}\}$. Finally, let $\Psi$ be such that $\mathfrak{P}_\Psi=\mathfrak{P}'$. It is clear that this reduction can be accomplished in logspace.

We claim $\Phi \in \QCSP(\mathcal{B})$ iff $\Psi \in \QCSP(\mathcal{C})$. (Forwards.) It is easy to see that exactly the same witnesses in $\mathcal{B}$ for the existential variables of $\Phi$ may be used in $\mathcal{C}$ for the existential variables in $\Psi$. This is because, whenever a universal variable of $\Psi$ is evaluated to $c$, then anything in $B \subset C$ witnesses any $\sigma$-relation. (Backwards.) Suppose that $\Psi \in\QCSP(\mathcal{C})$. We know that the first (universally) quantified variable of $\Psi$ corresponds to $a$, and therefore the remainder of $\Psi$ is true when this variable is evaluated anywhere in $B \subset C$. It follows that all the existential variables of $\Psi$ must be now be evaluated in $B \subset C$ and these will provide the witnesses in $\mathcal{B}$ for $\Phi$.

($\QCSP(\mathcal{C}) \leq_\mathsf{L} \QCSP(\mathcal{B})$.) Given an input $\Phi$ for $\QCSP(\mathcal{C})$ we build an input $\Psi$ for $\QCSP(\mathcal{B})$ in the following fashion. In $\mathfrak{P}_\Phi$ we look at all directed $F$-paths originating from elements in $\mathcal{A}_{P_\Phi}$. We remove all elements in $\mathcal{E}_{P_\Phi}$ other than those on such paths and derive $\mathfrak{P}'\subseteq \mathfrak{P}_\Phi$. Suppose $\Phi'$ is s.t. $\mathfrak{P}_{\Phi'}=\mathfrak{P}'$. We first claim that $\Phi \in \QCSP(\mathcal{C})$ iff $\Phi' \in \QCSP(\mathcal{C})$. This is easy to see as all the existential variables removed may be evaluated to $c$. It is also possible to compute $\Phi'$ from $\Phi$ in logspace, though for this one needs the result of \cite{ReingoldJACM}. Now, we look in $\mathfrak{P}_{\Phi'}=\mathfrak{P}'$ to see if there are any directed $F$-paths from elements in $\mathcal{A}_{P_\Phi}$ to other (not necessarily distinct) elements in $\mathcal{A}_{P_\Phi}$. If such a path exists, then $\Phi' \notin \QCSP(\mathcal{C})$ and we set $\Psi$ to be a fixed no-instance of $\QCSP(\mathcal{B})$ (say the sentence $\bot$). Otherwise, let $\mathfrak{P}''$ be the $\sigma$-partitioned structure that is the $\sigma$-reduction of the $\sigma \uplus \{F\}$-partitioned structure $\mathfrak{P}'$, and let $\Psi$ be s.t. $\mathfrak{P}_{\Psi}=\mathfrak{P}''$ (again note that we can compute $\Psi$ from $\Phi'$ in logspace).

We claim $\Phi' \in\QCSP(\mathcal{C})$ iff $\Psi \in\QCSP(\mathcal{B})$. (Forwards.) Since $\Phi'$ is true on $\mathcal{C}$ then $\Phi'$ is true on $\mathcal{C}$ when all universal variables are evaluated to $B \subset C$. In this case all existential variables of $\Phi'$ must also be evaluated to $B \subset C$ and it follows that these witnesses will also suffice for $\Psi$ on $\mathcal{B}$. (Backwards.) It is clear that the existential witnesses for $\Psi$ on $\mathcal{B}$ will work also for $\Phi'$ on $\mathcal{C}$.
\end{pf}

\section{Further work}

In \cite{LaroseLotenTardifJournal}, a characterisation of those CSPs that are first-order expressible is given. Might there be a characterisation of QCSPs that are first-order expressible? One problem is that our method of encoding positive Horn sentences uses a potentially infinite signature. It is not hard to see how a linear order on the variables might allow this to be made finite (highest in the order is outermost universal; next is existential etc.). 

Another question involves finding other classes (than $c$-valid) that are a microcosm for QCSPs. Feder and Vardi addressed this question for CSPs, finding, e.g,, that digraphs form such a microcosm \cite{FederVardi} (as they also do for QCSPs).

\bibliographystyle{acm}
\bibliography{she-short}

\end{document}